\documentclass[]{spie}  

\usepackage{amsmath,amsfonts,amssymb}
\usepackage{graphicx}
\usepackage[colorlinks=true, allcolors=blue]{hyperref}

\title{CAPD: A Context-Aware, Policy-Driven Framework for Secure and Resilient IoBT Operations}

\author[a]{Sai Sree Laya Chukkapalli}
\author[a]{Anupam Joshi}
\author[a]{Tim Finin}
\author[b]{Robert F. Erbacher}
\affil[a]{University of Maryland, Baltimore County, Baltimore, MD 21250}
\affil[b]{United States Army Research Laboratory,
Adelphi, MD 20783}

\begin{document} 
\maketitle

\begin{abstract}
The Internet of Battlefield Things (IoBT) will advance the operational effectiveness of infantry units. However, this requires autonomous assets such as sensors, drones, combat equipment, and uncrewed vehicles to collaborate, securely share information, and be resilient to adversary attacks in contested multi-domain operations. CAPD addresses this problem by providing a context-aware, policy-driven framework supporting data and knowledge exchange among autonomous entities in a battlespace.  We propose an IoBT ontology that facilitates controlled information sharing to enable semantic interoperability between systems. Its key contributions include providing a knowledge graph with a shared semantic schema, integration with background knowledge, efficient mechanisms for enforcing data consistency and drawing inferences, and supporting attribute-based access control. The sensors in the IoBT provide data that create populated knowledge graphs based on the ontology. 

This paper describes using CAPD to detect and mitigate adversary actions. CAPD enables situational awareness using reasoning over the sensed data and SPARQL queries.  For example, adversaries can cause sensor failure or hijacking and disrupt the tactical networks to degrade video surveillance. In such instances, CAPD uses an ontology-based reasoner to see how alternative approaches can still support the mission. Depending on bandwidth availability, the reasoner initiates the creation of a reduced frame rate grayscale video by active transcoding or transmits only still images.
This ability to reason over the mission sensed environment, and attack context permits the autonomous IoBT system to exhibit resilience in contested conditions. 
\end{abstract}


\section{INTRODUCTION}
\label{sec:intro}  
Rapid advancements in Internet of Things (IoT) applications have seen this technology used in several domains ranging from homes to industries, vehicles, and hospitals. These advances have also led the military towards the adoption of IoT based autonomous systems for the battlespace Ref.~\citenum{rapid}. This concept, described as the Internet of Battlefield Things Ref.~\citenum{russell18,cameron3internet}, integrates a network of  on device \& wearable sensors, actuators, and both ground and aerial semi autonomous vehicles to enhance information exchange across infantry units. However, sensors deployed in the battlespace are susceptible to a broad spectrum of attacks and often experience harsh environmental conditions leading to failure or disruption of services. These cyberattacks can cause massive damage to the battle plans as an adversary can control or disrupt the sensors to mislead the commanders or the AI controlling the autonomous units. So we need to design solutions to identify attacks, defend their targets, mitigate their risks, and exhibit resilience and graceful degradation. 

We have developed a context-aware, policy-driven framework designed to support attribute-based access control and incorporating resilience strategies through reasoning over sensed data to tackle attacks by the adversary. 
The contextual information for drawing inferences is gathered by collecting and continuously monitoring device data. When an adversary attacks sensors, they fail to send data or provide incorrect data. To overcome this drawback, defining context-aware policies to reason over information generated by the sensors identifies the type of attack on the sensor and further mitigates the risk by running defensive techniques. We view the elements of the battlefield IoBT system for multi domain operations (MDO) as autonomous agents that interact with one another to obtain the overall situational picture Ref.~\citenum{finin2002agents}, collaborating under constraints identified by policies Ref.~\citenum{toninelli2005rule}. These approaches build on our prior work that detected attacks in mobile {\em ad-hoc} networks Ref.~\citenum{parker2004intrusion}. In addition, the policy engine can also help infer alternative strategies for information exchange among the assets under adversarial attacks. 

The key contributions in this paper are:
\begin{itemize}
    \item We create a high level ontology for IoBT and MDO by creating new ontologies and building on existing ones to capture assets, people, access control attributes, mission context and situational awareness. 

    \item We populate a knowledge graph located in the policy engine by linking knowledge generated by sources with the IoBT ontology. The policy engine queries the knowledge graph to detect attacks and come up with mitigation strategies to ensure mission objectives.
    
    \item We demonstrate our approach using  a small testbed that mimics a real life scenario for a scouting platoon. 
\end{itemize}

The rest of the paper is organized as follows: Section \ref{sec:relatedwork} contains related work. Section \ref{sec:approach} explains the components of context-aware policy-driven framework, while Section \ref{sec:usecase} describes a use case scenario to demonstrate resilience under attack. Finally, we conclude and discuss the ongoing work in Section \ref{sec:conc}.

\section{Related Work}
\label{sec:relatedwork}
The advantages of incorporating policies grounded in access control models as explained in Ref.~\citenum{sandhu1994access} to secure and preserve the privacy of the user's data across multiple sectors are well known. Such techniques play a huge role in controlling access or information exchange between devices deployed in an IoT environment. Access control models such as Discretionary Access Control (DAC) Ref.~\citenum{DAC}, Mandatory Access Control (MAC) Ref.~\citenum{sandhu1994access}, and Role-Based Access Control (RBAC) Ref.~\citenum{sandhu1998role} models have been developed and used in the past. In DAC, the owner determines the user's access to information based on their identity. However, this model is limited as copying the information between objects cannot be controlled and it is also susceptible to exploitation through unauthorized access to sensitive information. The MAC model overcomes these limitations by providing a stricter control. This was essentially achieved by assigning security labels to both users and objects. RBAC is comparatively a more flexible and administrative-friendly access control model than DAC and MAC based on fixed and predetermined policies. The downside of the RBAC model is the role-permission explosion problem, where too many permissions assigned to roles make it hard to keep track.

More recently, the Attribute-Based Access Control (ABAC) model proposed by Jin et al. in Ref.~\citenum{jin2012unified} has gained momentum as it addresses the limitations presented by the above models. This model employs user and object attributes (or characteristics) to authorize decisions for operations by a subject (e.g., users and processes)  on the objects (like databases or files) in a system. Moreover, access control models, when combined with Web Ontology Language (OWL) Ref.~\citenum{mcguinness2004owl} provide flexibility in writing fine-grained context-aware policies. In previous work Ref.~\citenum{ROWLBAC} we showed how OWL could be used to implement and extend the standard RBAC model, providing a more expressive way to describe policies.

This contributes to the flexibility of access control decisions by reasoning in distributed multi-agent environments. Work done by Li et al. Ref.~\citenum{li2010securing} on securing and preserving the privacy of patients' health records exploited fine-grained access control policies for preventing multiple users from accessing sensitive information.

Similarly, Xue et al. Ref.~\citenum{xue2019attribute} show us how  attribute-based collaborative access control policies help in providing access control for public cloud storage. Multiple sectors such as smart homes Ref.~\citenum{sikder2020kratos, dutta2020context, yahyazadeh2019expat}, smart spaces Ref.~\citenum{hosseinzadeh2016semantic, pasquale2017topology}, smart farms Ref.~\citenum{chukkapalli2020ontologies, chukkapalli2020smart}, smart fisheries Ref.~\citenum{chukkapalli2021ontology}, 
smart grids Ref.~\citenum{suciu2021sealedgrid, li2015enabling, ruland2018firewall},
healthcare Ref.~\citenum{barhoun2019extended, figueroa2019attribute, li2021efficient} have also developed and applied ontology-based access control like ABAC framework for securing their ecosystems. In this paper, we aim to secure the constrained battlefield environment by supporting operations across a network of sensors and military units. In addition, we define context-aware access control policies and reason over them to detect and mitigate adversaries' attacks.

\section{Context-aware, policy-driven framework}
\label{sec:approach}

The Internet of Battlefield Things enables a range of heterogeneous automated army assets to exchange information across infantry units to support real-time operations in a battlefield. However, since they operate in contested environments, IoBT systems should expect to be attacked. In order to surmount such attacks it is necessary for the IoBT to exhibit resilience. 
Our approach to support resilience in battlefield involves integration of data from IoBT systems and sensors with the ontologies to create knowledge graphs. These are used to make decisions about how to respond when a system is attacked by reasoning over these knowledge graphs and background information such as mission objectives. We describe below the three main components of our framework such as context gathering, internet of battlefield things ontology and knowledge graph population and reasoning component.

\subsection{Context Gathering}

The goal of the CAPD framework is to gather information from assets present on the battlefield to exhibit resilience in an attack. To achieve this, we create an IoBT testbed having a range of heterogeneous automated army assets that mimic the battlefield environment in real-time. These automated army assets provide contextual information about their surroundings, such as the device's location, operations performed, network-level bandwidth, IP address, time, etc.

For example, drones send a continuous video feed to the commander in control in the battlespace. The video feed provides details about assets spotted at a particular time, location, and ongoing activities. These facts generated are shared with permitted assets to derive more contextual facts. All the information gathered is stored in the knowledge base and utilized by the policy engine for making a decision.

\subsection{Internet of Battlefield Things Ontology} \label{ontology}

\begin{figure} [ht]
   \begin{center}
   \begin{tabular}{c} 
   \includegraphics[height=9cm]{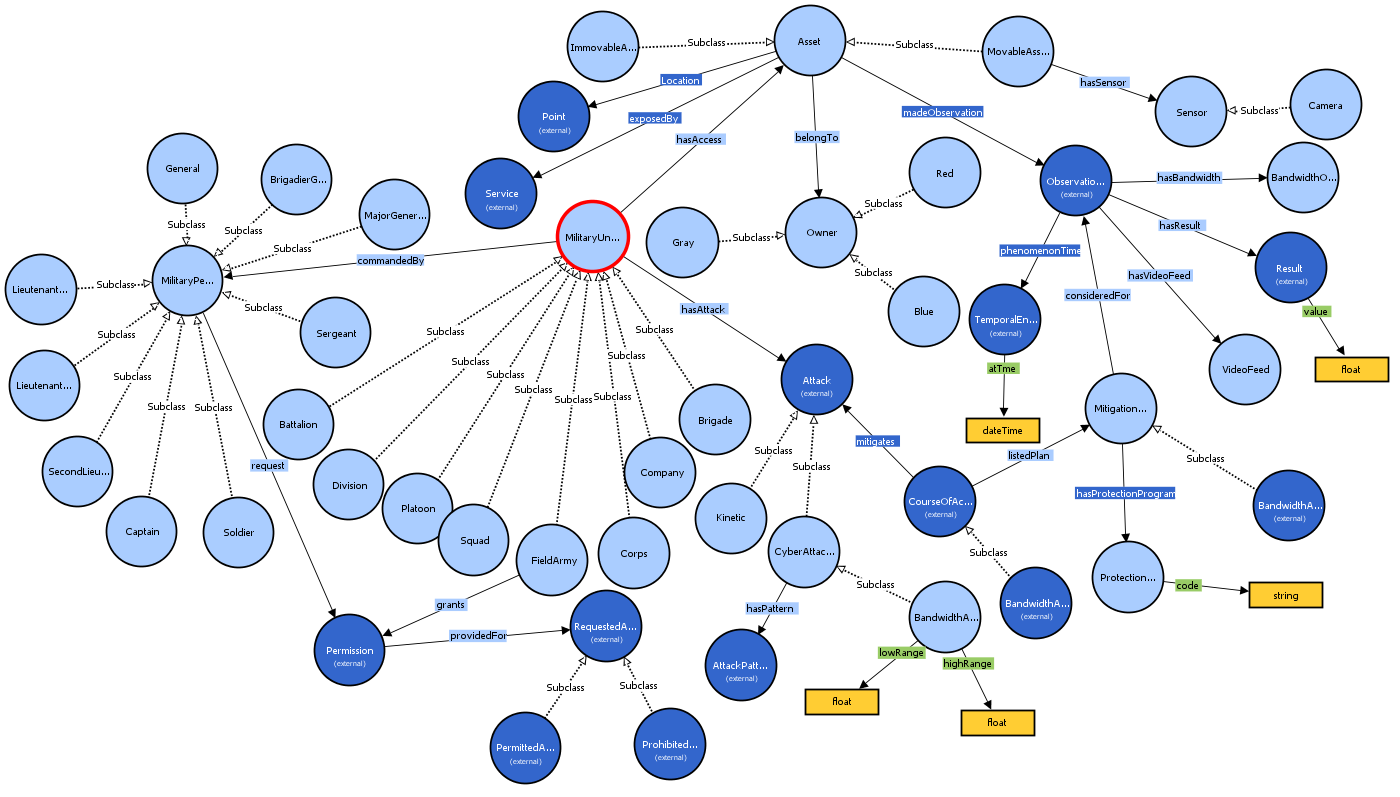}
   \end{tabular}
   \end{center}
   \caption[example] 
   { \label{fig:iobt} 
Internet of Battlefield Things ontology.}
   \end{figure}

We developed a semantically rich ontology for securing autonomous networks and systems including unmanned air and ground vehicles to enhance situational awareness in the battlefield. Our ontology promotes knowledge exchange between agents in contested environments by re-using multiple classes and properties from other commonly used ontologies. Figure \ref{fig:iobt} depicts various components of our ontology that capture attack-related scenarios from the Unified Cybersecurity Ontology (UCO) Ref.~\citenum{syed2016uco} and attribute-based access control (ABAC) concepts Ref.~\citenum{sharma2016representing}. We also add concepts representing kinds of resilience to the ontology and leverage the existing W3C Sensor IoT-Lite Ref.~\citenum{bermudez2016iot} and Geospatial Ref.~\citenum{budak2006geospatial} ontologies to represent interactions on the battlefield.

Using RDF-based knowledge graphs enables the use a variety of open-source tools developed by the community as well as commercial tools.  For example, SHACL (Shapes Constraint Language) Ref.~\citenum{shacl} is a W3C standard that is used to declare constraints on the content, structure and meaning of a knowledge graph that can be automatically checked to identify incorrect or missing data. There are many reasoning tools, both open-source and commercial, that can be used to infer and add new nodes and relations to the graph based on its initial structure.

Since the IoBT is a very dynamic environment some information that was once true, such as the location of a vehicle, will become false and must be changed.  Data must also be removed or changed, if we determine that it reported by an edge device that we've determined to be unreliable or compromised.  We plan to modify one of the open-source SWRL reasoning tools Ref.~\citenum{swrl} to add a truth-maintenance capability based on our earlier work Ref.~\citenum{finin1989tms}.  This will automatically remove inferred changes to the knowledge graph that are no longer supported as nodes or relations are removed. It can also be used to generate explanations for inferred relations based on the asserted facts and rules used to infer them.

An ontology for RDF-based knowledge graphs includes a collection of \textit{classes} for concepts that are organized into a taxonomy via \textit{subClassOf} relations and a collection of \textit{properties} forming a taxonomy using the \textit{subPropertyOf} relation. Properties typically have constraints on their domain and range, i.e., the classes or data types they can connect.  The classes and properties are used to define \textit{instances} of one or more classes with a set of properties and values. The OWL language provides a very rich set of additional constraints that can be expressed about classes and properties. 

Our current IOBT ontology can be viewed via its permanent url Ref.~\citenum{iobt_ontology}. Some of the important classes and properties of our ontology are explained below.

\subsubsection{Classes}
Here, we define the fundamental concepts for the battlefield in the form of classes. For instance, the Sensor class characterizes all kinds of sensors present on the battlefield.

\begin{itemize}
     \item \textit{Asset} class: This class represents the assets owned by the military to perform various operations on the battlefield. It has two subclasses such as \textit{MovableAsset} and \textit{ImmovableAsset}. The \textit{MovableAsset} class represents the movable sensors such as drones, unmanned ground vehicle (UGV), unmanned aerial vehicle (UAV), etc. and \textit{ImmovableAsset} class indicates a fixed sensor like unattended ground sensors (UGS).
     
     \item \textit{Attack} class: This class represents types of attacks that happen in the battlefield. It has two subclasses named \textit{Kinetic} and \textit{CyberAttack}.
     
     \item \textit{Observation} class: This class provides us with data points recorded by the sensors that belong to the \textit{Asset} class.
    
    \item \textit{BandwidthObservation} class: This class represents the network bandwidth of automated army assets present on the battlefield.
    
    \item \textit{TemporalEntity} class: This class represents the temporal information of the recorded observations generated by the sensors. 
    
    \item \textit{Result} class: This class provides us with the data points generated by the assets present on the battlefield.
    
    \item \textit{BandwidthAttack} class: This class is a subclass of \textit{CyberAttack}. It provides information
   about the attempts of an adversary to cripple the communication channels' bandwidth and thus prevent information exchange. 
   
   \item \textit{CourseOfAction} class: This class characterizes the measures taken to defend an attack from happening or reduce the impact of an attack.
    
    \item \textit{MitigationPlan} class: This class provides us with a wide variety of mitigation techniques based on the type of an attack to circumvent the threats caused by adversaries across all connected units.
    
    \item \textit{ProtectionProgram} class: Every mitigation plan involves a series of steps to reduce risk caused by an adversary. This class describes the methods for each of the mitigation techniques.
\end{itemize}

\subsubsection{Relations}
Relations establish a link between classes. We describe below how our classes relate to each other. 

\begin{itemize}
    
    \item \textit{hasBandwidth}: Represents the link between 
    \textit{Observation} class and \textit{BandwidthObservation} class. This property aids in identifying network bandwidth levels for each of the sensors that belong to the \textit{Asset}class.
    
    \textbf{Domain}: Observation
    
    \textbf{Range}: BandwidthObservation

    \item \textit{phenomenonTime}: This relation establishes a link between \textit{Observation} class and \textit{TemporalEntity} class indicating time at which network bandwidth is recorded.
    
    \textbf{Domain}: Observation
    
    \textbf{Range}: TemporalEntity

     \item \textit{hasResult}: This relation determines the  value recorded by various individual assets at a particular timestamp.
    
    \textbf{Domain}: Observation
    
    \textbf{Range}: Result
    
    \item \textit{mitigates}: Represents link between \textit{CourseOfAction} class and \textit{Attack} class.
    It provides information on action to be taken when attacked by an adversary.

    \textbf{Domain}: CourseOfAction
    
    \textbf{Range}: Attack

    \item \textit{listedPlan}: This relation between \textit{CourseOfAction} class and \textit{MitigationPlan} class determines mitigation technique to be considered based on the detected attack found exploiting the army assets.
    
    \textbf{Domain}: CourseOfAction
    
    \textbf{Range}: MitigationPlan
    
   \item \textit{hasProtectionProgram}: Relationship where the subject entity belongs to the \textit{MitigationPlan} class and object entity belongs to the \textit{ProtectionProgram} class indicates protection program code designed for each mitigation technique.
    
    \textbf{Domain}: MitigationPlan
    
    \textbf{Range}: ProtectionProgram

\end{itemize}
    
\subsection{Knowledge Graph Population and Reasoning}

Knowledge graphs play a vital role in encapsulating the domain knowledge from an ontology linked with data generated from sensors. They aid in a more profound understanding of the contextual information to make decisions. In our work, we use semantic web technologies such as Resource Descriptions Framework (RDF) Ref.~\citenum{rdf} and SPARQL Protocol and RDF Query Language Ref.~\citenum{SPARQL} to populate and query a knowledge graph. Here the entities and their relationships as described in Section \ref{ontology} serve as a schema for the knowledge graph.

We incorporate a policy engine that populates and queries the knowledge graph to support reasoning over contextual information integrated with the semantic schema of the battlefield. The populated knowledge graph in RDF format is present in the policy engine, where the asset information is in a structured format suitable for reasoning and drawing inferences. Further, the policy engine queries the knowledge graph utilizing SPARQL to infer a protection program for mitigating the adversaries' impact caused by an attack.

\section{Use case scenarios}
\label{sec:usecase}
Our CAPD framework identifies attacks on the battlefield and uses context-driven policies to defend against them. To demonstrate the capabilities of our framework, we build an upper-level ontology to describe the ``battlefield” and link with other underlying ontologies for sensors, cybersecurity, and access control as described in Section \ref{ontology}. First, we added military assets and their communication and interactions to the semantic schema. We also included resilience concepts related to the mission. In our setup, we utilize low power edge devices like the NVIDIA Jetson Nano platform Ref.~\citenum{JetsonNano20} to implement our mission and show how our ontology can be utilized to create populated knowledge graphs from sensed data and reason over them in response to attack scenarios. 

\textbf{Use Case 1}: Consider a scenario in the battlefield where an adversary attacks to significantly reduce network bandwidth of a military asset named \textit{Asset\_A} (Camera) that is continuously transferring video feed to another asset named \textit{Asset\_B} (Lieutenant's handheld device). In this case, we monitor the network bandwidth level between \textit{Asset\_A} and \textit{Asset\_B} associated with \textit{BandwidthObservation} class and identify whether the observed value falls under low, medium, or high bandwidth categories based on ranges determined for each category. The intuition is that an adversary may try to jam or degrade the connection.

Based on the category of network bandwidth, the reasoner infers a protection program to facilitate information exchange in the contested environment. An example of a SPARQL query for the above scenario is presented below:
 
\begin{verbatim}
    PREFIX bf:<http://purl.org/ArtIAMAS/battlefield#>
    PREFIX sosa: <http://www.w3.org/ns/sosa/phenomenonTime#>
    PREFIX stix: <http://purl.org/cyber/stix/mitigates#>

    SELECT (?TS as ?Time) (?BA as ?BandwidthStage) (?code AS ?Mitigation_Program)
    WHERE {
      ?BA a bf:BandwidthAttack;
          bf:lowRange ?l;
          bf:highRange ?h .
      ?BO a bf:BandwidthObservation;
          sosa:phenomenonTime  ?TS;
          bf:hasResult  ?Res .
      ?Res bf:value ?val .
      FILTER (?val >= ?l && ?val <= ?h).
      ?BAM stix:mitigates ?BA;
           bf:listedPlan ?MitigationPlan.
      ?MitigationPlan stix:hasProtectionProgram ?PP .
      ?PP bf:code ?code .
    }
    ORDER BY ?TS
\end{verbatim}
\noindent
The above query looks for the range of network bandwidth 
availability in order to decide whether to communicate the regular resolution color video, gray-scale video or still pictures. If the network bandwidth is high, color video is transferred. For medium level bandwidth gray scale video is transferred and pictures for lower bandwidth. If the network bandwidth is very low, we run an image detection algorithm locally on \textit{Asset\_A} to compute only the number of tanks spotted and transfer the information to the commander in control.

\textbf{Use Case 2}: The adversary jams network connection between assets to prevent the exchange of information. For example, \textit{Asset\_A} utilizes a fourth-generation (4G) connection by default to support the data exchange over a range upstream. However, the CAPD framework identifies the link as jammed by the adversary and switches the network connection from 4G technology to Long Range Wide Area Network (LoRaWAN). Since LoRaWAN consumes less power while covering a wide area with a lower data transmission rate, we also need to make changes in the content transmitted similar to use case 1.

\textbf{Use Case 3}: An adversary completely subverts the operations of camera, by physical covering it, or causing smoke/dust etc. In this case, our policy framework automatically switches to gathering information from the microphone and shares the information with the lieutenant. For example, instead of sending the video of advancing armor, we could listen to the sound to estimate whether the enemy armor was in the region. 

\textbf{Use Case 4}: The adversary orchestrates an attack on assets at a particular location. In this situation, our CAPD framework identifies movable assets (drones) in the nearby area and transmits a command for repositioning their video focus. As a result, the commander in control receives information to understand the neighboring plot's situation better. Similarly, we can utilize our framework to investigate new attacks and create policies to capture adaptation under attacks.


\section{conclusion}
\label{sec:conc}

Resilience on the battlefield is imperative to accommodate failures of sensors or disruption of services caused by an adversary. We have developed the context-aware, policy-driven (CAPD) framework to facilitate information exchange and demonstrate resilience under attacks. We use information generated by military assets and our Internet of Battlefield Things (IoBT) ontology to populate a knowledge graph. Further, policy engine queries populated knowledge graphs to infer knowledge that supports secure operations on the battlefield. We also demonstrated the ability of our CAPD framework to reason over the context in contested environment with the help of a use case scenario. In our ongoing work, we plan to extend our framework by investigating how adversaries may poison sensor data and identifying corresponding resilience strategies.

\acknowledgments 
This research was supported by U.S. Army Grant No. W911NF2120076. The authors would like to thank Dr. Roberto Yus for the suggestions provided during the development of our ontology.

\bibliography{report} 
\bibliographystyle{spiebib} 

\end{document}